\documentclass[twocolumn]{aastex63}
\shorttitle{A Suggested Final VLA Configuration}
\shortauthors{Wrobel \& Walker}

\begin{document}

\title{A Suggested Final Configuration for the Very Large Array}

\author[0000-0001-9720-7398]{J. M. Wrobel}
\affiliation{National Radio Astronomy Observatory, P.O. Box O,
  Socorro, NM 87801, USA}

\author[0000-0002-6710-6411]{R. C. Walker}
\affiliation{National Radio Astronomy Observatory, P.O. Box O,
  Socorro, NM 87801, USA}

\correspondingauthor{J. M. Wrobel}
\email{jwrobel@nrao.edu}

\received{2022 May 13 as ngVLA Memo \# 97 for consideration by the
  VLA/VLBA to ngVLA Transition Advisory Group}

\begin{abstract}
If the construction of the ngVLA begins in 2026, its sensitivity is
expected to match that of the VLA by late 2029. At that juncture it is
anticipated that open-skies observing will cease on the VLA and
commence on the ngVLA. We suggest that during 2026-2029 the VLA be
held in a customized final configuration encompassing portions of its
standard A, B, C and D configurations. Such a final VLA configuration
would (1) help minimize the cost of VLA operations and maximize the
pace of ngVLA construction and commissioning; (2) help VLA users pivot
to the high-resolution, high-frequency research topics that are
projected to headline the ngVLA science program; and (3) help mitigate
the effects of source confusion during responses to transients in the
era of the Rubin Observatory and LIGO A+.
\end{abstract}


\section{Context and Motivation}

The next-generation VLA (ngVLA) is envisaged to be an interferometric
array operating at frequencies between 1.2 and 116 GHz, with ten times
the sensitivity and angular resolution of the VLA and ALMA
\citep{mur18,sel18,mck19}. If the construction of the ngVLA begins in
2026, its sensitivity is expected to approximately match that of the
VLA by late 2029.\footnote{https://ngvla.nrao.edu/} At that juncture
it is expected that PI-driven, open-skies observing will cease on the
VLA and commence on the ngVLA as Early Science \citep{for19}. The NRAO
has begun working with the community to identify and evaluate possible
options for such a transition.\footnote{
https://science.nrao.edu/enews/15.5/} In the interim we are guided by
some draft concepts mentioned by \citet{cha19}, notably the
possibility that the VLA continue to operate at a reduced level during
2026-2029.

Here, we explore a hyothetical reduction in one capability of the VLA,
namely its reconfigurability. Specifially, we suggest that the VLA be
held in a customized final configuration encompassing portions of its
standard A, B, C and D configurations. Such a final VLA configuration
would (1) help minimize the cost of VLA operations and maximize the
pace of ngVLA construction and commissioning, by freeing staff from
VLA reconfiguration activities; (2) help VLA users pivot to the
high-resolution, high-frequency research topics that are projected to
headline the ngVLA science program \citep{mur18,wro20a,wro20b}; and
(3) help mitigate the effects of host-galaxy and cosmological
confusion during responses to transients in the era of the Rubin
Observatory and LIGO A+ \citep{ive19,rei19}.

\section{Approach and Results}

For each of its standard A, B, C and D configurations, the VLA offers
a power-law spacing of the nine antennas placed along each of its
three equiangular arms \citep{tho80,nap83}. The distance $d_n$ from
the center of the array of the $n^{th}$ antenna per arm, counting
outward from the center, is proportional to $n^{1.716}$. The different
standard configurations have different proportionality constants. The
values chosen for those constants offer two advantages. First, it
means that some antenna pads can be shared among the configurations,
so only 24 pads per arm are needed to accommodate all standard
configurations. Those pad locations are shown in Figure~1. Below, we
will make use of each arm's 24 standard pad identifers $p$ that span 1 to
72 with gaps \citep[see Table~1 in][]{tho80}. Second, the scaling
between the standard configurations resembles that between three of
the VLA's original observing bands, facilitating imaging at matched
angular resolutions among those bands. With the advent of complete
frequency coverage between 1 and 50 GHz on the VLA \citep{per11} and
robust data-weighting schemes \citep{bri95}, this second advantage has
become less significant.

We suggest that during 2026-2029 the VLA be held in a customized final
configuration encompassing portions of all its standard
configurations. We opt to avoid populating the two innermost pads per
arm, $p=1$ and $p=2$, as such short spacing information can be
obtained with single dish facilities. We also opt to avoid populating
the outermost pad per arm, $p=72$, as this will help reduce the
operational burden. This leaves us with a set of 21 pad locations per
arm that we wish to populate with nine antennas per arm. To do so, we
take two steps.

First, we seek a power-law spacing of the nine antennas per arm,
spread between the innermost antenna's $d_1 = 89.9$~m on pad $p=3$ and
the outermost antenna's $d_9 = 17160.8$~m on pad $p=64$. These
extremes define a power-law exponent $log_{10}(d_9/d_1) / log_{10}(9)
= 2.390$ and lead to the set of nine desired distances $d_n^{desired}$
given in Table~1.

Second, for simplicity we invoke the VLA's power-law model for its pad
locations per arm. We then search among pads $p=4, ..., 56$ per arm to
find the seven pads that come closest to achieving the desired
distances $d_n^{desired}$ for seven additional antennas. The seven
closest-pad identifiers plus the two end-defining pads are given in
Table~1, along with their closest-pad distances $d_n$.

\begin{deluxetable*}{cccccccccc}
\tablecolumns{10} \tablewidth{0pc}
\tablecaption{A Suggested Final Configuration for the VLA}
\tablehead{
  \colhead{(1)} & \colhead{(2)} & \colhead{(3)} & \colhead{(4)} &
  \colhead{(5)} & \colhead{(6)} & \colhead{(7)} & \colhead{(8)} & 
  \colhead{(9)} & \colhead{(10)}}
\startdata
Antenna index, $n$ & 1 & 2 & 3 & 4 & 5 & 6 & 7 & 8 & 9 \\
Desired distance, $d_n^{desired}$ (m) & 89.9 & 471.3 & 1242.1 &
          2470.4 & 4211.1 & 6511.1 & 9411.7  & 12950.2 & 17160.8 \\
Closest pad identifier, $p$ & 3 & 8 & 14 & 20 & 28 & 36 & 48 & 56 & 64 \\
Closest pad distance, $d_n$ (m) & 89.9 & 484.0 & 1264.4 & 
          2331.8 & 4153.9 & 6393.6 & 10474.7 & 13646.6 & 17160.8 \\
\enddata
\end{deluxetable*}

The closest-pad identifiers in Table~1 define our suggestion for the
VLA's final configuration. Armed with those identifiers, we use {\tt
  SCHED} to access their catalogued locations and generate $(u,v)$
coverage plots for short (0.2 hour) and long (8.0 hour) tracks,
subject to an antenna elevation limit of 15 degrees. The $(u,v)$
coverage plots are shown on A-configuration scales in Figures~2 and 3,
on B-configuration scales in Figures~4 and 5, on C-configuration
scales in Figures~6 and 7, and on D-configuration scales in Figures~8
and 9. These figures indicate reasonable $(u,v)$ coverage on the
standard scales long familiar to VLA users.

\section{Summary and Next Steps}

We explored a hypothetical reduction in one capability of the VLA,
namely its reconfigurability, during the years of a VLA-to-ngVLA
transition. We identified a power-law configuration for the VLA that
involves portions of its standard A, B, C and D configurations. We
suggested that the VLA be held in this final, customized configuration
during the transition years, and mentioned some operational and
scientific advantages of doing so.

A specific next step is to use simulations to study the performance
parameters and image fidelity of our suggested final configuration for
the VLA. We look forward to learning the community's reaction to our
suggestion. We also look forward to learning about the alternate ideas
that will emerge as the NRAO engages with community stakeholders to
identify and evaluate possible options for the VLA/VLBA-to-ngVLA
transition.

\begin{figure*}[!thb]
\centering
\includegraphics[scale=0.50]{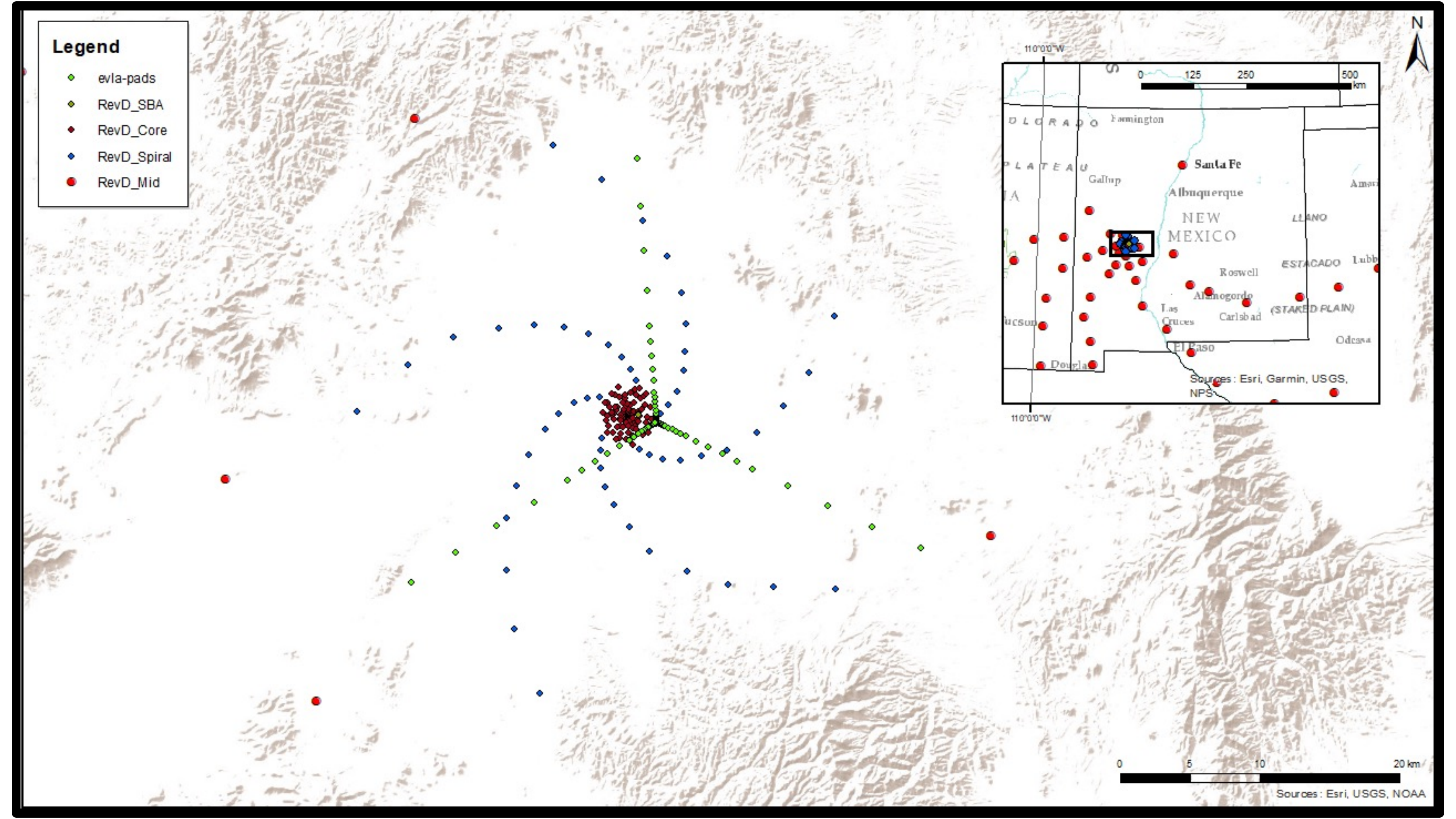}
\caption{Map of the Plains of San Agustin, New Mexico, USA. Green
  symbols mark the 24 standard pad locations on each arm of the VLA.
  Other symbols mark potential Rev D locations of ngVLA antennas.
  Terrain features are also indicated. Adapted from \citet{car22}.}
\end{figure*}

\acknowledgments We thank Joe Carilli for generating Figure~1. The
NRAO is a facility of the National Science Foundation (NSF), operated
under cooperative agreement by AUI. The ngVLA is a design and
development project of the NSF operated under cooperative agreement by
AUI.

\clearpage

\begin{figure*}[!t]
\centering
\includegraphics[angle=-90,scale=0.55]{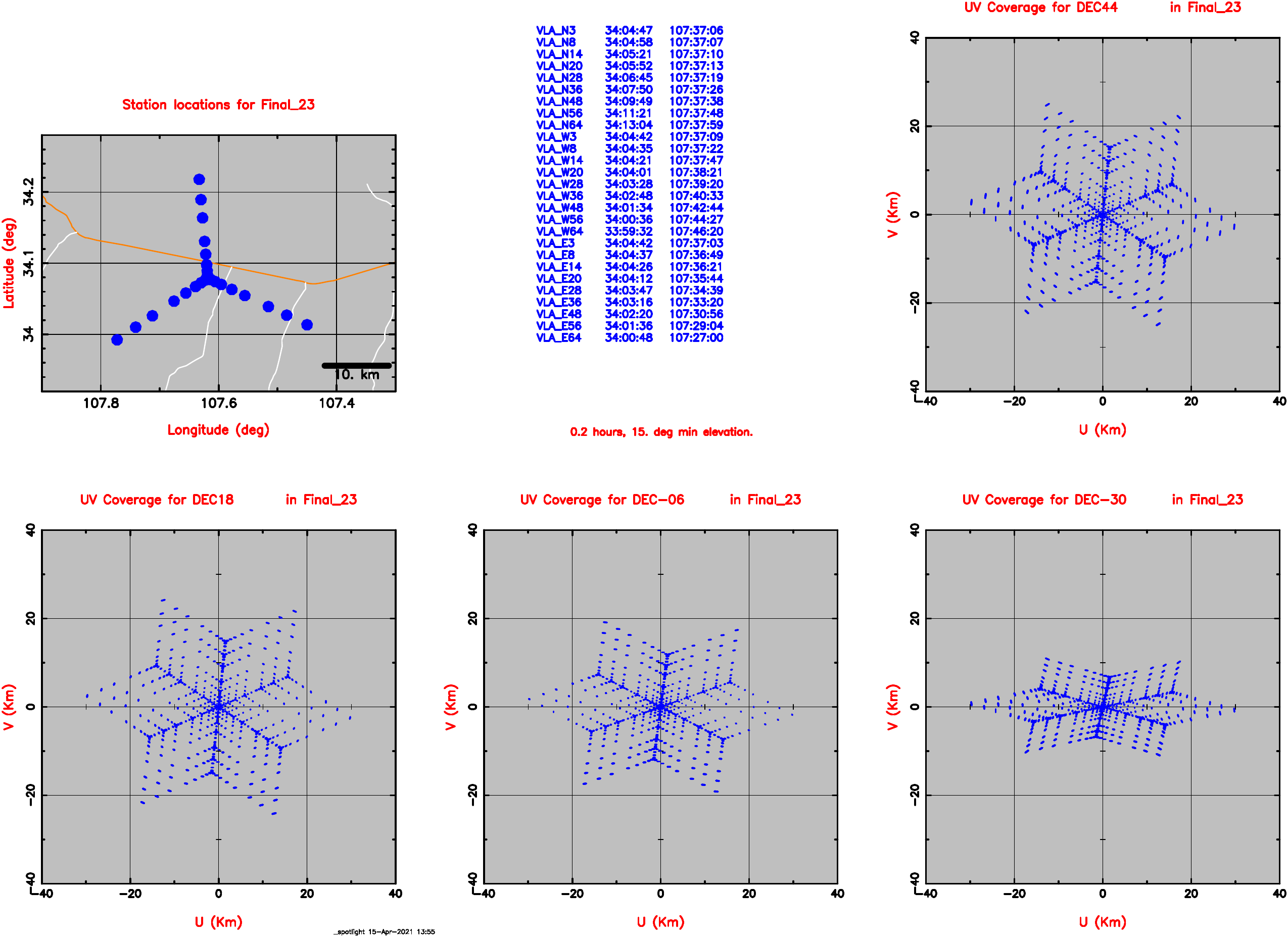}
\caption{$(u,v)$ coverage over $\pm$ 40~km. Short track.}
\end{figure*}

\begin{figure*}[!b]
\centering
\includegraphics[angle=-90,scale=0.55]{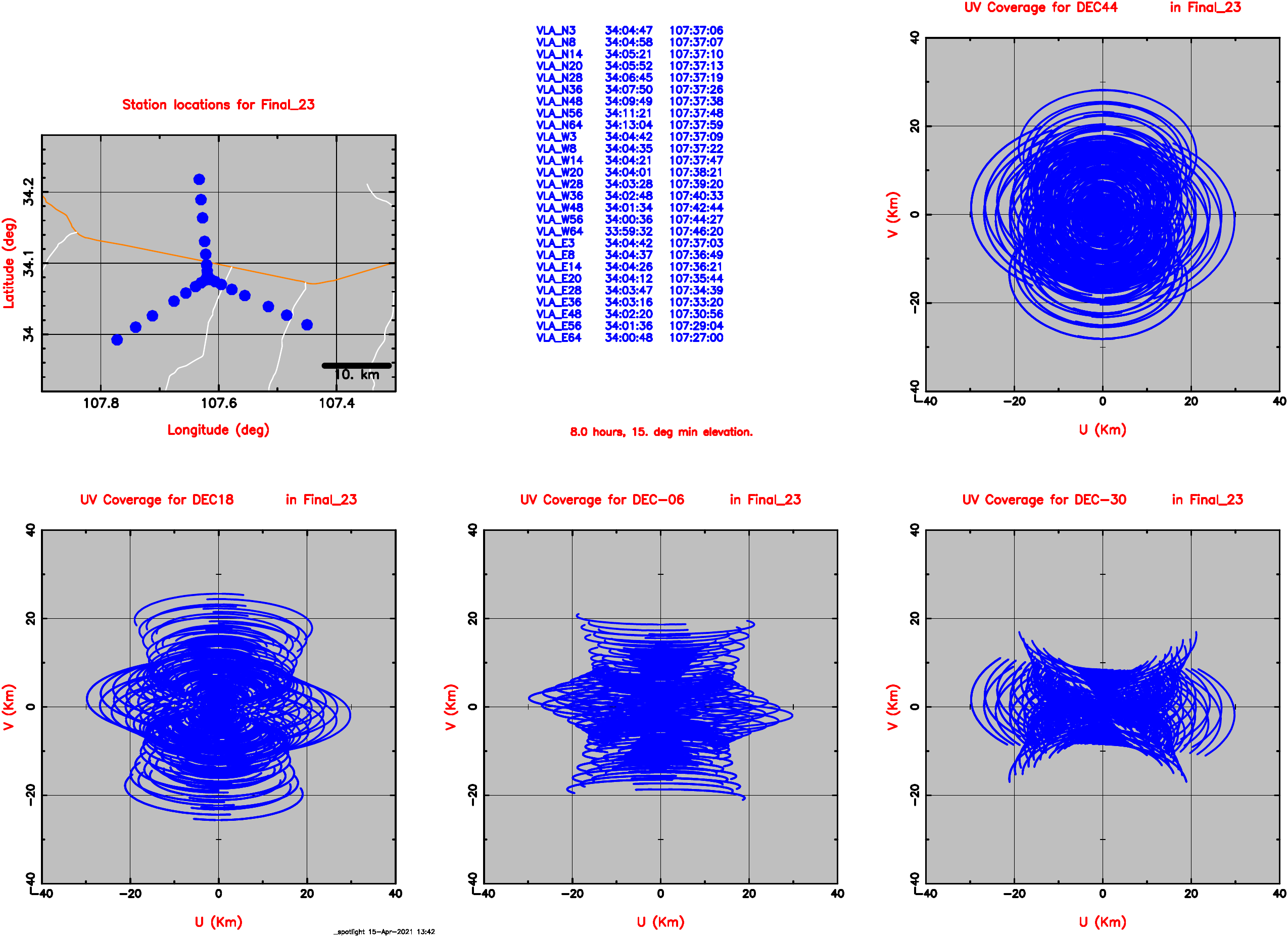}
\caption{$(u,v)$ coverage over $\pm$ 40~km. Long track.}
\end{figure*}

\clearpage

\begin{figure*}[!t]
\centering
\includegraphics[angle=-90,scale=0.55]{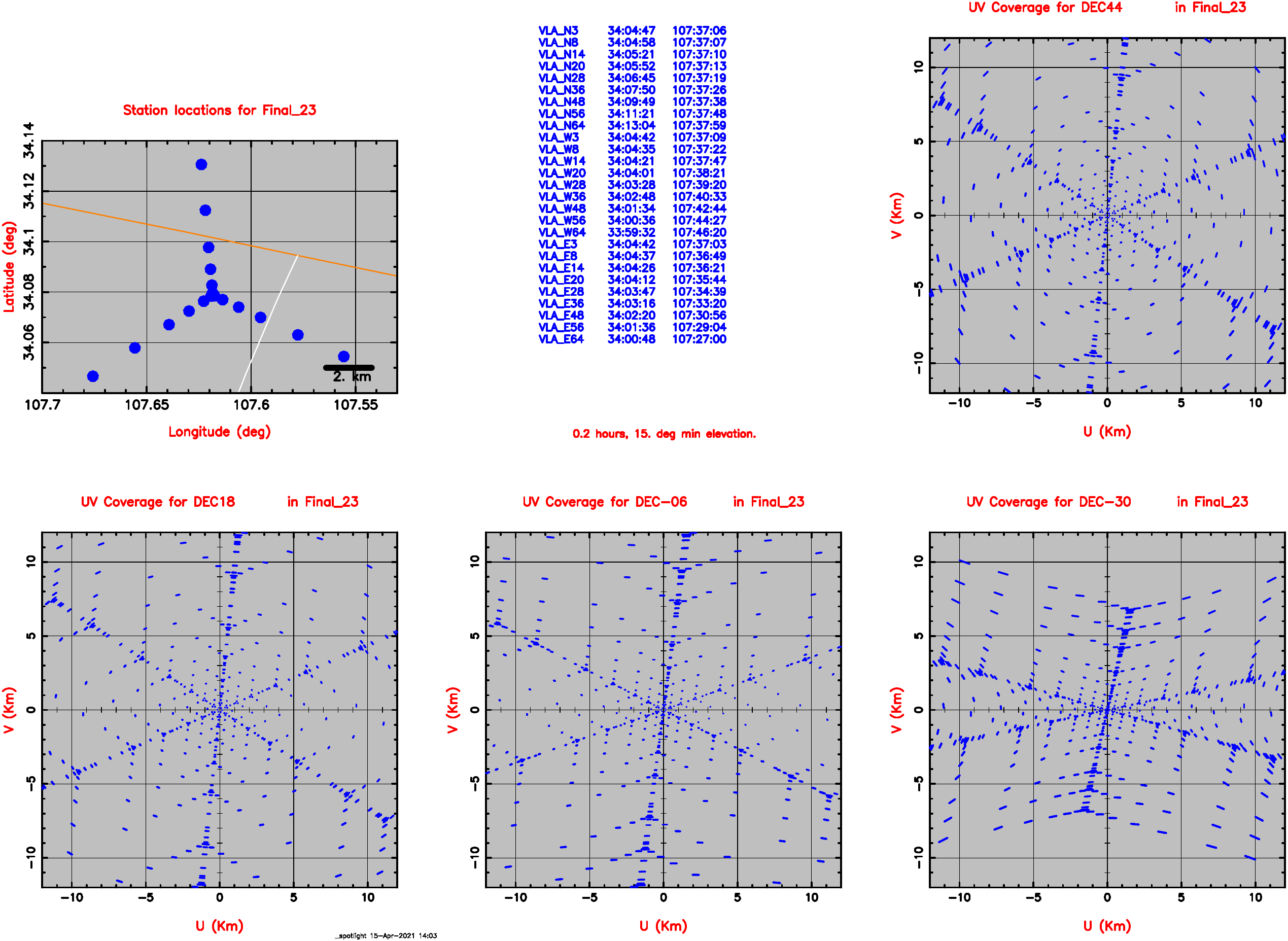}
\caption{$(u,v)$ coverage over $\pm$ 12~km. Short track.}
\end{figure*}

\begin{figure*}[!b]
\centering
\includegraphics[angle=-90,scale=0.55]{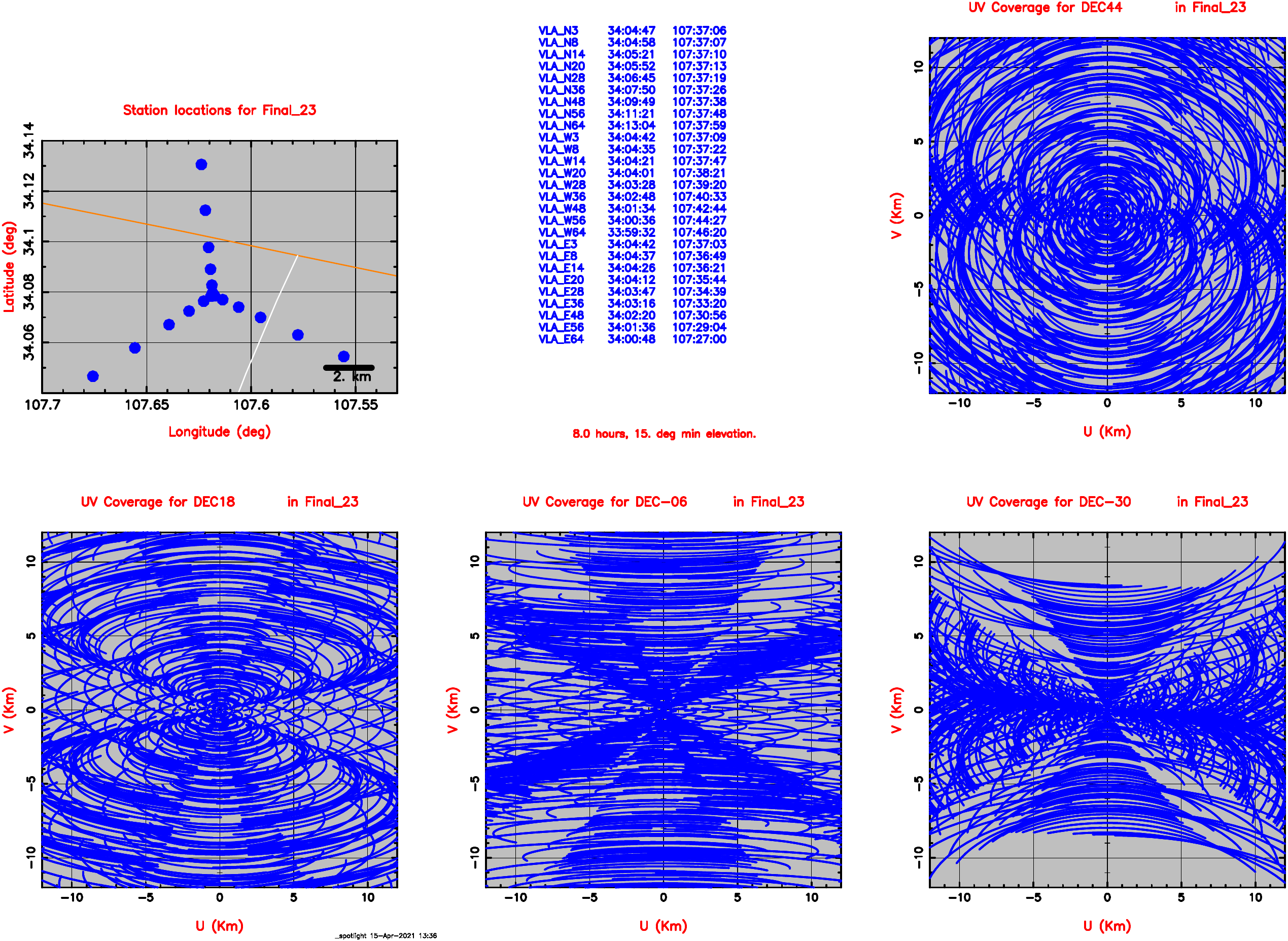}
\caption{$(u,v)$ coverage over $\pm$ 12~km. Long track.}
\end{figure*}

\clearpage

\begin{figure*}[!t]
\centering
\includegraphics[angle=-90,scale=0.55]{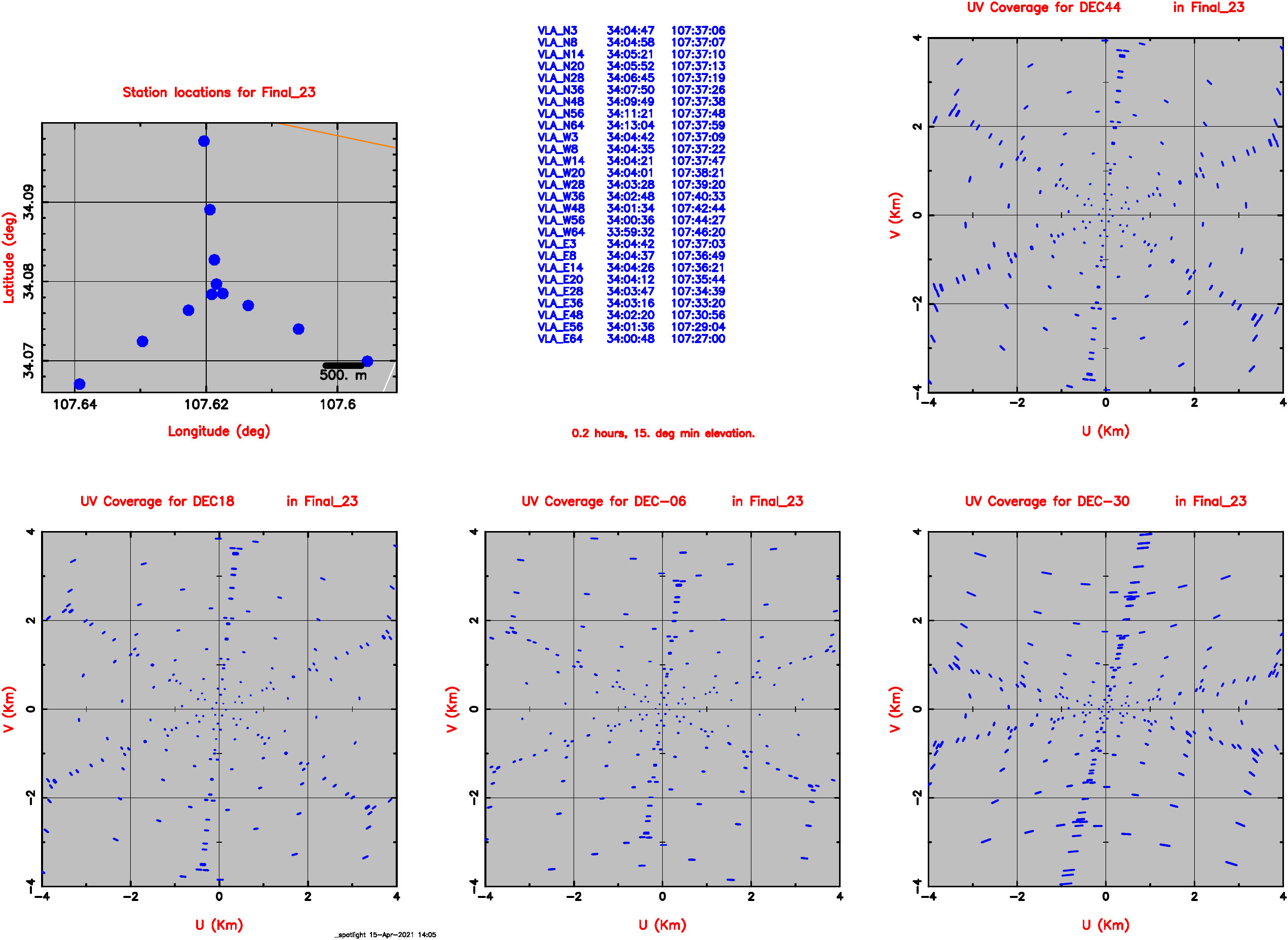}
\caption{$(u,v)$ coverage over $\pm$ 4~km. Short track.}
\end{figure*}

\begin{figure*}[!b]
\centering
\includegraphics[angle=-90,scale=0.55]{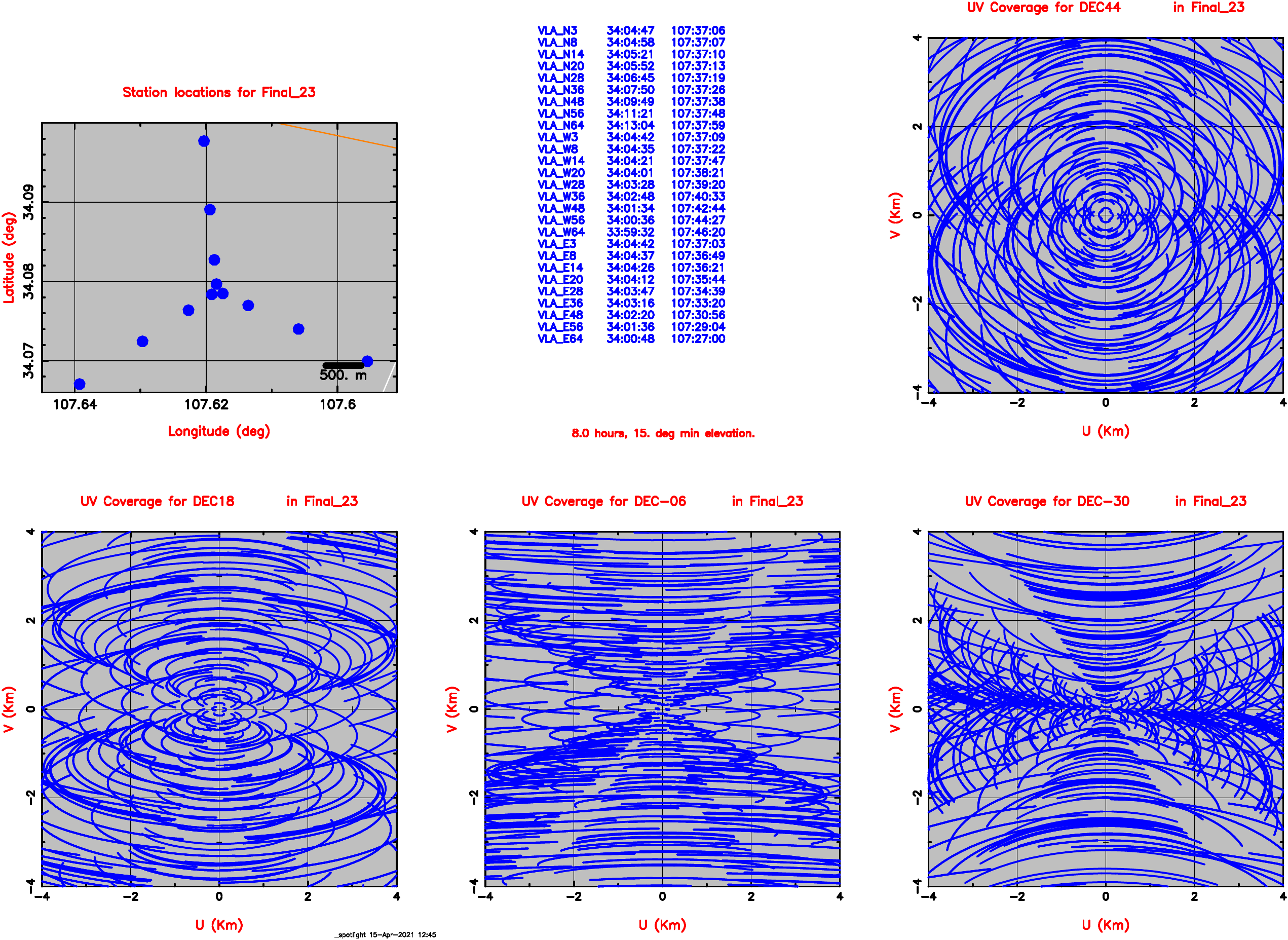}
\caption{$(u,v)$ coverage over $\pm$ 4~km. Long track.}
\end{figure*}

\clearpage

\begin{figure*}[!t]
\centering
\includegraphics[angle=-90,scale=0.55]{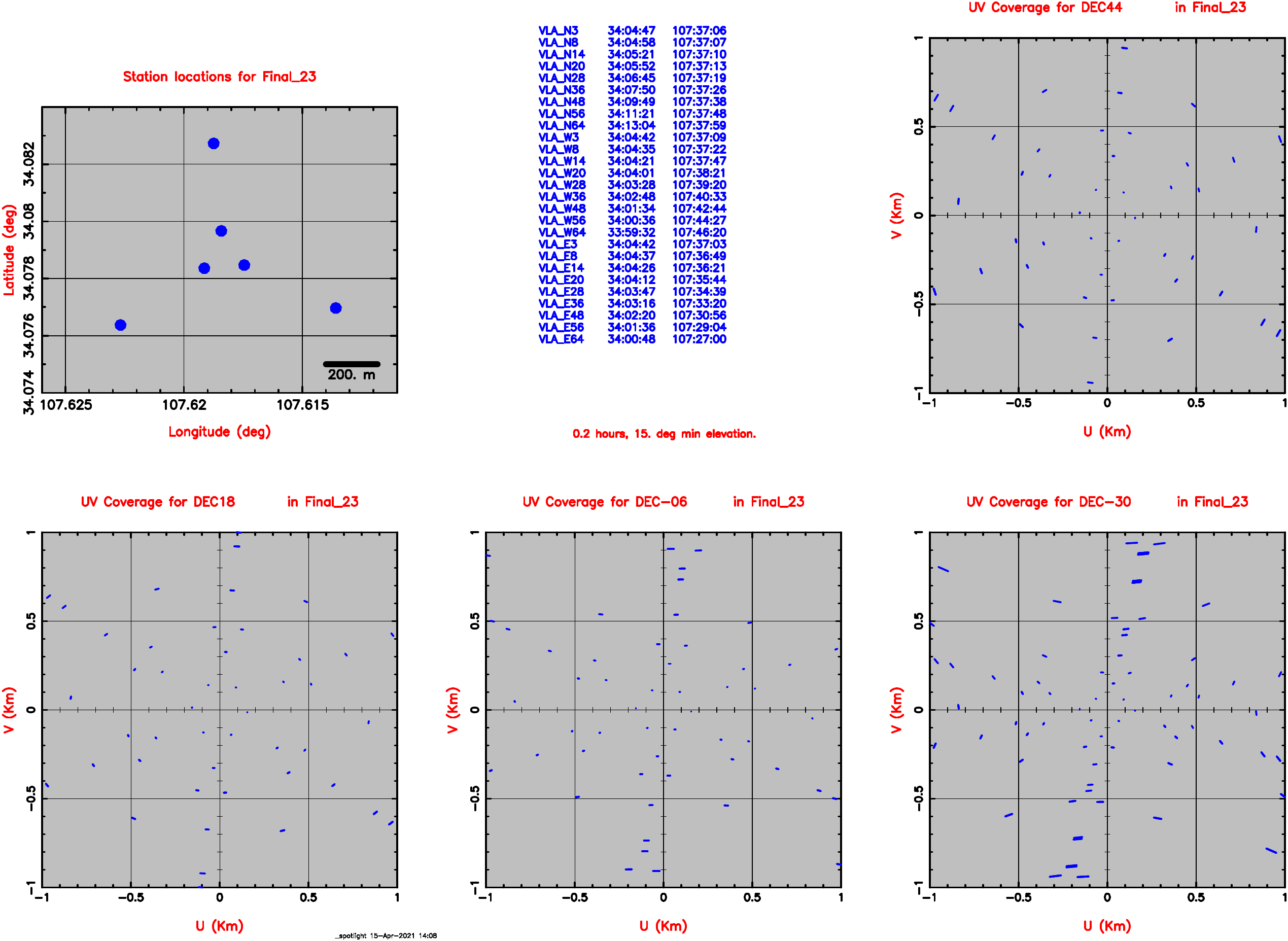}
\caption{$(u,v)$ coverage over $\pm$ 1~km. Short track.}
\end{figure*}

\begin{figure*}[!b]
\centering
\includegraphics[angle=-90,scale=0.55]{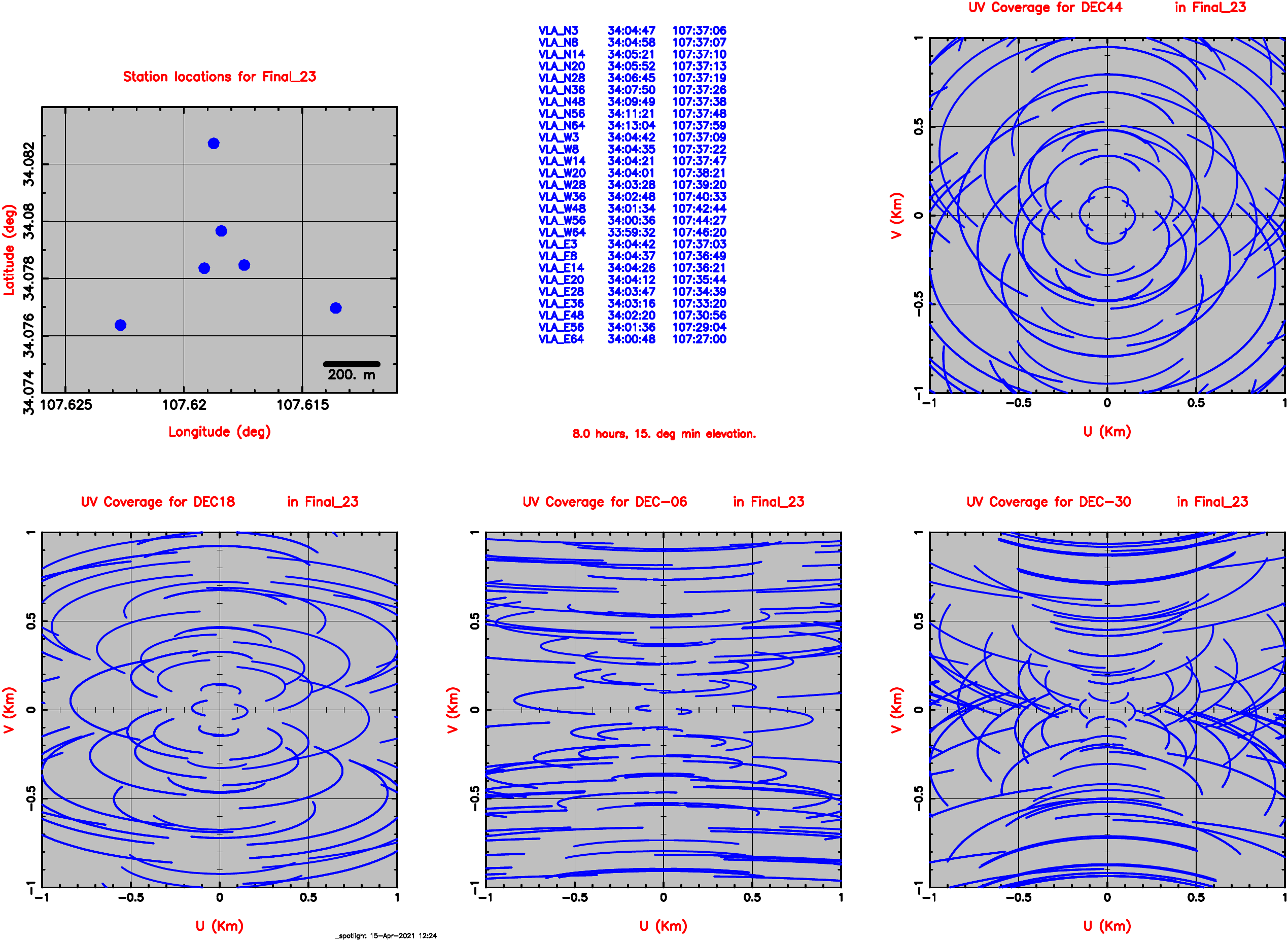}
\caption{$(u,v)$ coverage over $\pm$ 1~km. Long track.}
\end{figure*}

\clearpage

\end{document}